\documentclass[12pt,epsfig]{article}

\usepackage{graphicx}
\usepackage{subfig}
\usepackage{amsfonts}
\usepackage{amssymb}
\usepackage{amsmath}
\usepackage{mathdots}
\usepackage{placeins}
\usepackage{bbold}
\usepackage{hyperref}

\newskip\humongous \humongous=0pt plus 1000pt minus 1000pt
  \newif\ifdtup

\def\frac#1#2{ {{#1} \over {#2} }}

\def\VEV#1{\left\langle #1\right\rangle}
\def\eg{\hbox{\em e.g. }}
\def\ie{\hbox{\em i.e. }}

\def\beq{\begin{equation}}
\def\eeq{\end{equation}}

\newcommand{\beqa}{\begin{eqnarray}}
\newcommand{\eeqa}{\end{eqnarray}}
\newcommand{\ba}{\begin{array}}
\newcommand{\ea}{\end{array}}
\newcommand{\bmat}{\begin{pmatrix}}
\newcommand{\emat}{\end{pmatrix}}
\newcommand{\bcas}{\begin{cases}}
\newcommand{\ecas}{\end{cases}}

\def\Tr{\mbox{Tr}\;}

\begin{document}

\title
{Taylor expansions on Lefschetz thimbles \\
(and not only that)}

\author
{F.~Di~Renzo, S. Singh and K.~Zambello\\
\small{Dipartimento di Scienze Matematiche, Fisiche e Informatiche, Universit\`a di Parma} \\
\small{and INFN, Gruppo Collegato di Parma} \\
\small{I-43100 Parma, Italy}\\
}

\maketitle

\begin{abstract}
Thimble regularisation is a possible solution to the sign problem,
which is evaded by formulating quantum field theories 
on manifolds where the imaginary part of the action stays constant
(Lefschetz thimbles). A major obstacle is due to the fact that one 
in general needs to collect contributions coming from more than one 
thimble. 
Here we explore the idea of performing Taylor expansions on Lefschetz 
thimbles. We show that in some cases we can compute expansions in 
regions where only the dominant thimble contributes to the result in 
such a way that these (different, disjoint) regions can be
bridged. This can most effectively be done via Pad\'e approximants. 
In this way multi-thimble simulations can be circumvented. The 
approach can be trusted provided we can show that the analytic
continuation we are performing is a legitimate one, which thing we can
indeed show.  
We briefly discuss two prototypal computations, for which we obtained
a very good control on the analytical structure (and singularities) of the
results. All in all, the main strategy that we adopt is supposed to 
be valuable not only in
the thimble approach, which thing we finally discuss.
\end{abstract}

\section{Introduction: thimble regularisation and single thimble dominance}

Lattice regularisation provides an effective framework for a
non-perturbative definition of Quantum Field Theories. It also enables
numerical computations: in the euclidean formulation, a lattice-regularised QFT
resembles a statistical physics problem, the functional integral
defines a decent probability measure and Monte Carlo simulations
are viable. However, this is not always the case. When a complex
action is in place, we have no probability measure to start with and
there is no obvious way to set up a Monte Carlo scheme. This is
one consequence of what is 
known as the sign problem. Among other theories, QCD at non-zero chemical
potential is plagued by a sign problem and at the moment we have no 
effective way to tackle the investigation of its (supposedly rich) 
phase diagram by lattice methods (for a clean explanation of the sign problem in
the framework of QCD see \cite{SIGNproblemQCD}; for an up-to-date
account of the status of the efforts to constrain the QCD phase
diagram see \cite{OweLAT2019}). \\

\noindent
Thimble regularisation is an elegant (attempt at a) solution to the 
sign problem. Following seminal papers by Witten \cite{Witten1,Witten2}, it has 
been introduced with the idea that the sign problem can be beaten at
a fundamental level. Thimble regularisation is conceptually simple: one changes
the domain of integration by complexifying the degrees of freedom 
and defines the theory on manifolds where
the imaginary part of the action stays constant. These manifolds are
the so-called Lefschetz thimbles \cite{Aurora,Kikukawa}. In practice, the method has many
subtleties. A major one has to do with a fundamental feature, {\em i.e.} the so-called thimble
decomposition. While there are cases in which a single contribution
(attached to the so-called dominant thimble) accounts for the solution
of the problem at hand, in general a given quantity is computed summing contributions
attached to many thimbles. This feature in turn has to do with the
occurrence of Stokes phenomena. The latters take place at given points
in the space of the parameters which define the theory: for certain
values of the parameters there is no meaningful thimble
decomposition. While {\em at those points} there is no thimble
decomposition, {\em across those points} there is a discontinuity in
thimble decompositions, which nevertheless does not generally imply
a discontinuity in physical results. This basic feature will play a
major role in the following. \\

\noindent
In the original formulation \cite{Aurora} a single thimble dominance
conjecture was put forward. The underlying idea is very simple: one
defines the theory as the functional integral restricted to the 
dominant thimble, {\em i.e.} the one associated to the absolute
minimum of the real action. From very general (semiclassical)
arguments, this contribution is expected to be more and more enhanced 
in the thermodynamic limit. At a more fundamental level, the 
regularisation of a field theory on the dominant thimble defines a 
local quantum field theory with exactly 
the same symmetries, the same number of degrees of freedom (belonging 
to the same representations of the symmetry groups) and the same local 
interactions as the original theory. Moreover, the perturbative
expansion on the dominant thimble is exactly the same computed in 
standard perturbation theory in the original formulation. Quite
interestingly, in the case of 
the relativistic Bose gas this approximation proved to work very
well \cite{BoseGAS}.
The Bose gas was of course only one success. Needless to say, if 
the dominant thimble dominance hypothesis held true
for a wide range of theories, that would be a major success: numerical
simulations for thimble regularisation would be in the end not that difficult. 
The Thirring model was shown to be a (first) counterexample: in this case the 
dominant thimble does not capture the (complete) correct result
\cite{ThirringKiku,PauloAndrei1}. This was a major motivation for
the exploration of alternative formulations somehow inspired by thimbles. 
The idea of deforming the original domain of integration
is indeed a very general one (from this point of view the complexification of the degrees of
freedom is quite an obvious thing to do).
Alternatives to thimble regularisation were put forward, {\em e.g.} the holomorphic
flow \cite{PauloAndrei1} or various definitions of 
sign-optimised manifolds, possibly selected by deep-learning techniques 
\cite{PauloAndrei2,PauloAndrei3,Mori}; for a recent, nice review of
most of these ideas see \cite{PauloAndreiReview}. \\

\noindent
In this work we want to explore the idea of performing Taylor
expansions on Lefschetz thimbles. All in all, the main idea is to compute
Taylor expansions in different regions of the parameter space of a
given theory, namely around points where only the dominant thimble
contributes to the result one is interested in. This could seem
somehow a lucky scenario, but we argue that that this can quite often
be the case. Through multiple expansions these (different, disjoint)
regions can be bridged and in this way multi-thimbles simulations can
be circumvented. This bridging can most effectively be obtained via
Pad\'e approximants. The question then should be: ``Can we trust this
bridging?'' The answer is yes if we can prove that we are going
through a legitimate analytic continuation. Needless to say, working
with Taylor expansions we are aiming at a control on analytic
contributions to the result and we will be blind to any non-perturbative
effect in the expansion parameter\footnote{We stress that the
  non-perturbative effects we are talking about are not the ones in
  the coupling constant, \ie the ones we are most often concerned with.}. 
Most noticeably, in the simple examples we preliminarily discuss in this
work, we show that we can have a very good control on the analytic
structure (and singularities) of the results. We stress that this good
control is
coming from multiple Taylor expansions at different points (in the end,
different values for the chemical potential) and Pad\'e approximants. 
While all this is
discussed in the framework of (and motivated by) thimble regularisation, 
it is important to recognise in what we do an overall strategy 
that does not apply only to this regularisation. This is in a sense a strategy
which can go well beyond thimbles. \\

\noindent
The paper is organised in a such a way that a minimal prior knowledge
of the subject is assumed. In section \ref{sec:ThimbleBASICS} we
collect the basics of thimble regularisation; in particular, we define
what a thimble decomposition is and we mention a couple of approaches
to multiple thimbles simulations we have been testing in the last few
years; the main content of the section is the focus on Stokes phenomena. 
Section \ref{sec:TaylorOnThimbles} contains the basic
description of the computational method we propose together with a
discussion of preliminary results for a couple of
theories, which (while admittedly simple) are supposed to be valuable
prototypes. Conclusions \ref{sec:concludendo}
are meant to recognise to which extent the
strategy we discuss can go beyond the application to thimble regularisation.

\section{Thimble decomposition and Stokes phenomena: a story of
  continuity and discontinuities} 
\label{sec:ThimbleBASICS}

\subsection{Basics of thimble decomposition and basic multiple
  thimbles computations}
\label{sec:multiThimble}

Let us summarise the general problem by writing 
\begin{subequations}
\label{eq:allVEV}
\begin{align}
\left\langle O \right\rangle \, & = \, Z^{-1} \, \int dx \; e^{-S(x)} \, O(x) \label{eq:basicComputation} \\
       & = \, \frac{\sum_{\sigma} \; n_{\sigma} \,
  e^{-i\,S_I(p_{\sigma})} \, \int_{\mathcal{J}_\sigma} dz \;
  e^{-S_R(z)}\; O(z)\; e^{i\,\omega(z)}}{\sum_{\sigma} \; n_{\sigma} \, e^{-i\,S_I(p_{\sigma})}\, \int_{\mathcal{J}_\sigma} dz \;
  e^{-S_R(z)}\; e^{i\,\omega(z)}} \label{eq:ThimbleDecomposition} \\
       & = \, \frac{\sum_{\sigma} \; n_{\sigma} \,
  e^{-i\,S_I(p_{\sigma})} \, Z_{\sigma} \left\langle O\,e^{i\,\omega} \right\rangle_{\sigma}
         }{\sum_{\sigma} \; n_{\sigma} \, e^{-i\,S_I(p_{\sigma})}\, \,
         Z_{\sigma} \left\langle e^{i\,\omega} \right\rangle_{\sigma}} \label{eq:ThimbleDecompositionBis}
\end{align}
\end{subequations}
It is assumed that $S(x) = S_R(x) + iS_I(x)$. $x$
stands collectively for the {\em real} degrees of freedom of our
original problem and $z$ for the degrees of freedom that are the
{\em complexification} of the latter. (\ref{eq:basicComputation}) is
the original formulation of the problem, while
(\ref{eq:ThimbleDecomposition}) is the thimble decomposition as
expected from Lefschetz/Picard theory. $p_{\sigma}$ are critical 
points where $\partial_z S = 0$. The thimbles $\mathcal{J}_\sigma$
attached to each critical point are the union of all
  the Steepest Ascent paths (SA) which are the solutions of 
$$\frac{d}{dt} z_i = \frac{\partial
  \bar{S}}{\partial \bar{z_i}}$$ 
stemming from a given critical point (initial condition). Both numerator
and denominator ({\em i.e.} the partition function) of 
(\ref{eq:basicComputation}) are rewritten as a linear combination of
integrals computed on the thimbles attached to critical points; 
the sum is formally extended to
all of them, but the coefficients  $n_{\sigma}$ can be zero for possibly many
critical points. Actually $n_{\sigma}=0$ for a critical point when the
associated {\em unstable} thimble (the union of
the Steepest Descent paths stemming from the critical point) does 
not intersect the original integration manifold. While the (constant) 
phase $e^{-i\,S_I(p_{\sigma})}$ is factored in front of each integral,
yet another phase enters the integrands. This is the so-called 
{\em residual phase} \cite{ResPhase} ($e^{i\,\omega}$) which accounts for the
orientation of the thimbles with respect to the
embedding manifold\footnote{Thimbles are manifolds of the same (real) dimension of the
original manifold the theory was formulated on, but they are embedded in
a manifold of twice that dimension.}.\\
In (\ref{eq:ThimbleDecompositionBis}) $\left\langle O  \right\rangle$ is
rewritten by defining
$$\left\langle X  \right\rangle_{\sigma} \, \equiv \, \frac{\int_{\mathcal{J}_\sigma} dz \;
  e^{-S_R}\; X}{\int_{\mathcal{J}_\sigma} dz \;
  e^{-S_R}} \, \equiv \, \frac{\int_{\mathcal{J}_\sigma} dz \;
  e^{-S_R}\; X}{Z_{\sigma}}.$$
Stated in this way, the thimble decomposition is a linear
combination of expectation values computed on single thimbles, with
coefficients proportional to the $Z_{\sigma}$. As a result,
multiple thimbles simulations amount to a given prescription for
obtaining, in one way or another, (a) the contribution attached 
to each given thimble $\left\langle O  \right\rangle_{\sigma}$
contributing to the result and (b)
the relative weights in (\ref{eq:ThimbleDecompositionBis}). Actually,
(b) turns out to be a harder task than (a). Despite the difficulties,
there are cases in which we could attempt and succeed in multiple
thimbles simulations. While these are admittedly preliminary steps in
an interesting direction, they are worth mentioning, if only to
appreciate the peculiar circumstances under which they worked out.\\

There are cases in which not only it turns out that non-null 
contributions come from a limited number 
of thimbles, but also a few of the latter are
related to each other due to symmetries. One possible strategy in
such cases is the one which was successful for QCD in $0+1$ 
dimensions \cite{QCD01}. In that case (due to a symmetry which is in
place) the correct result was obtained by taking into account only two
contributions according to
\begin{equation}
\VEV{O} \, = \, \frac{\VEV{O\,e^{i\,\omega}}_{\sigma_1} \,+\, \alpha
  \, \VEV{O\,e^{i\,\omega}}_{\sigma_2}}
{\VEV{e^{i\,\omega}}_{\sigma_1} \,+\, \alpha \, \VEV{e^{i\,\omega}}_{\sigma_2}}.
\label{eq:ThimbleDecompAlpha}
\end{equation}
(\ref{eq:ThimbleDecompAlpha}) is yet another rewriting of the thimble
decomposition. All in all, all our ignorance of relative weights is in
such a case coded in one single parameter, \ie $\alpha$. The value of
the latter can be fixed assuming one known measurement as a 
normalisation point. We can then predict the value of other
observables. It is obvious that in such a way we give up the hope of a
first principles derivation of relative weights. This is very much in
the spirit of general frameworks for (non-perturbative)
renormalisation. Quite interestingly, this appears to be possible also
in the framework of the Thirring model \cite{Kevin}. \\

It should be stressed that relative weights are quite easy to obtain
in a semiclassical approximation, which is also referred to as the
{\em gaussian approximation}. This suggests another strategy: one
starts with the relative weights as computed in such an approximation
and then compute corrections as the simulations proceeds. Once again,
this is not expected to work efficiently in every case. A case of
success was a minimal version of 
the so-called Heavy Dense approximation for QCD \cite{HDqcd}. Yet
another proposal for ``reweighting Lefschetz Thimbles'' was put forward
in \cite{reweightHeidelberg}.

\subsection{Deeper into the problem: Stokes phenomena}
\label{sec:Stokes}

Multiple thimbles simulations are a hard problem and
solutions so far have been admittedly a partial success. 
Our goal here is to find an alternative to them which goes beyond
the naive single thimble prescription. In order to 
proceed, let's have a look at Stokes phenomena: these
control the basic mechanism of the thimble decomposition. 
The main lesson to take home is that a thimble  
decomposition is never given once and forever. 
In the following we provide a simplified, informal discussion. The
interested reader is strongly referred to \cite{StudyThirringKiku} for
a nice discussion of the subject in the context of the Thirring model.\\

Loosely speaking, we have a thimble decomposition when the union
of a given number of thimbles is a convenient deformation of the
original domain of integration, just like
in standard applications of the Cauchy theorem. It is clear that the 
deformation provided by thimbles is not the only possible one 
(this is \eg the spirit of
\cite{PauloAndrei1,PauloAndrei2,PauloAndrei3,Mori}). 
Strictly speaking, thimbles provide a basis of the 
relative homology group which the integration cycle we are interested 
in belongs to. It is a very convenient basis because the imaginary
part of the action stays constant on thimbles. It is also a very
convenient basis because the coefficients in the linear combination
reconstructing a given path (the $n_{\sigma}$) are integers. Moreover
(as already said) we have a criterion to establish which thimbles do
not enter a decomposition: $n_{\sigma}=0$ whenever the unstable
thimble associated to a given critical point does not intersect the 
original domain of integration\footnote{It can be shown that the
  $n_{\sigma}$ have the meaning of intersection numbers.}. 
This has an important consequence, which is quite clear pictorially.
As they are different solutions of the same (first order)
differential equation subject to different initial conditions, 
{\em different thimbles can not cross each other}. Put in a
simple-minded (but maybe effective) way, they act as {\em barriers to each
other}: when the union of a given number of thimbles provide a {\em
correct} decomposition of the original integration contour, 
other thimbles are simply kept out.

\begin{figure}[ht]
\begin{center}
  \begin{tabular}{ccc} 
 \hspace{-0.5cm}
     \includegraphics[height=4.5cm,clip=true]{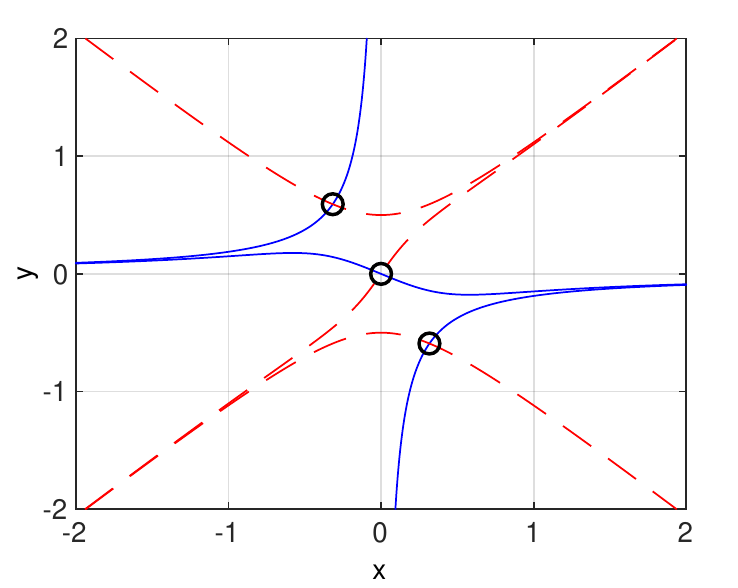}
     &
 \hspace{-0.5cm}
     \includegraphics[height=4.5cm,clip=true]{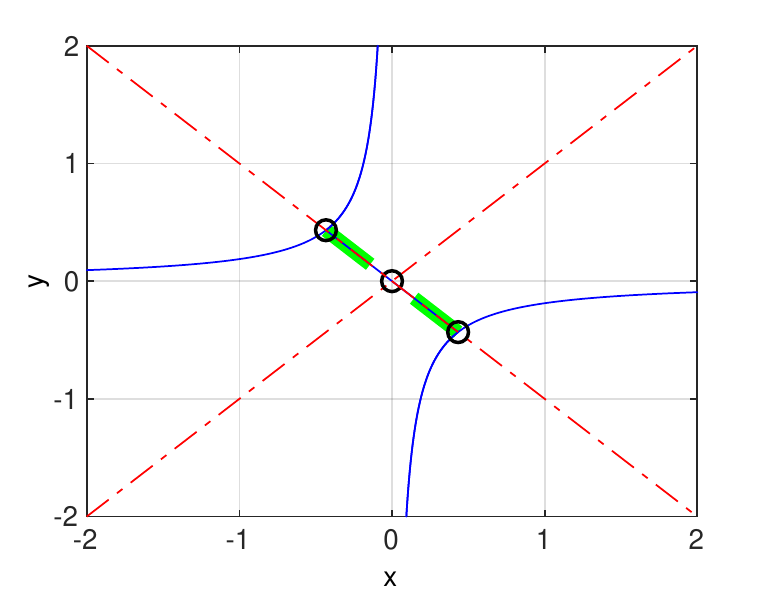}
     &
 \hspace{-0.5cm}
     \includegraphics[height=4.5cm,clip=true]{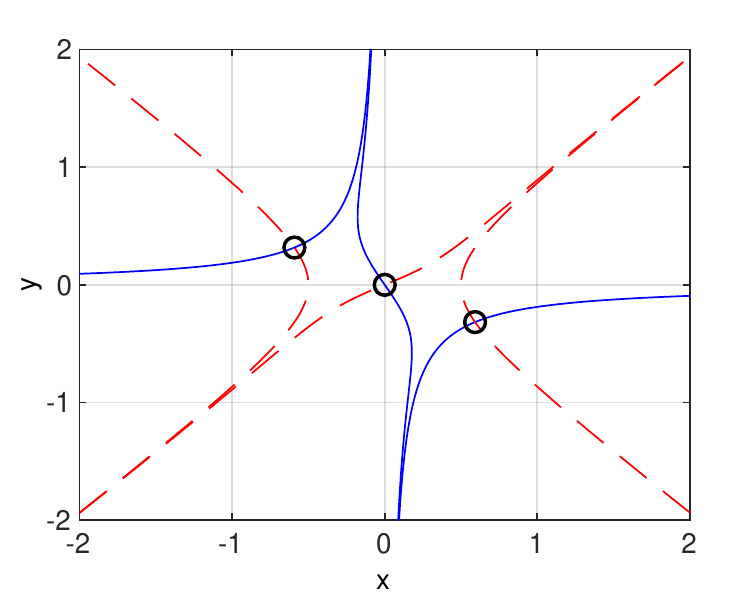}
  \end{tabular}
\end{center}
  \caption{Thimbles structure for a $0$-dim $\phi^4$ toy model:
    continuous (blue) lines are stable thimbles; dashed (red) lines
    are unstable thimbles. The three
    panels refer to different points in the parameter space of the
    theory. In the middle, an example of a Stokes phenomenon. For
    further details, see \cite{thimbleRMT}.}
  \label{fig:phi40d}
\end{figure}

In order to gain some insight, in Figure \ref{fig:phi40d} we plot what 
the problem looks like in a simple toy model, \ie the $0$-dim $\phi^4$
theory (the problem amounts to computing a simple real
integral; see \cite{thimbleRMT}). In the left panel, we can see the
correct thimble decomposition at a given point of the parameter space 
of the theory; stable thimbles are
depicted as continuous (blue) lines: as it is manifest, one single thimble is
enough to get a correct deformation of the original domain of
integration (the latter being the real axis)\footnote{Note how only one 
unstable thimble intersects the original domain of integration; 
unstable thimbles are depicted as dashed (red) lines.}.
Probing different points in the parameter space, one finds that critical points
and associated thimbles do move around in the manifold embedding the 
original one (\ie the complexified manifold). As they smoothly move
around, they are always subject to the constraint of not 
crossing each other. Thus the thimbles that contribute to the
decomposition of the original domain of integration keep on keeping
the others out. \\
There is only one way thimbles can cross each other: we need
two thimbles to sit on top of each other. This means that two different
critical points are connected by a SA/SD path, \ie the 
stable thimble of one sits on top of the unstable thimble of the
other. When this occurs we are in presence of a Stokes phenomenon: see
the central panel. As it is clear from the figure, {\em at a Stokes 
phenomenon the thimble decomposition fails}. \\
After a Stokes phenomenon has occurred, the {\em relative} arrangement 
of thimbles can change and a different thimble decomposition is in
place (see right panel).\\

The toy model we referred to is  a trivial example: one
can recognise the occurrence of a Stokes phenomenon by direct
inspection. One could then think that tracking Stokes phenomena could
be an almost impossible task in a typical case. This is not the case,
since we have a clear signal that a Stokes phenomenon {\em can} occur:
when two critical points are connected as described above, the
imaginary part of the action takes the same value.
To look for Stokes phenomena, we change the values of the parameters
describing the theory; let's denote collectively these parameters
by $\xi$.
%\footnote{In the case of the Thirring model there
%is one single parameter $\frac{\mu}{m}$.}. 
As the $\xi$ vary, a critical point
$p_{\sigma}$  moves around and the value of the imaginary 
part of the action associated to it (and to the stable 
and unstable thimbles attached to it) describes a curve 
$S_I^{(\sigma)}(\xi)$. In order that a Stokes phenomenon occurs, 
two curves need to intersect, \ie $S_I^{(\sigma)}(\xi_0)=S_I^{(\sigma')}(\xi_0)$. For a
beautiful description of the procedure which we just sketched, we
refer the interested reader to \cite{StudyThirringKiku}.

We end this sketchy discussion with a trivial, but in practice relevant 
observation. Reconstructing the correct thimble decomposition is not
necessarily the end of story. It can well be that one (or more)
thimble(s) entering the correct thimble decomposition is (are) so damped with
respect to other contributions that its (their) contribution is {\em
  de facto} negligible. There are cases in which this is evident from
the semi-classical approximation: this has to do with the 
$e^{-S_R(p_{\sigma})}$ that can be factored in front of the integral
associated to a critical point $p_{\sigma}$ and which was one of the
main rationales for the single thimble dominance hypothesis.

\subsection{Discontinuities vs continuity}

Stokes phenomena mark {\em discontinuities in the thimble
decomposition} and one is left with the problem of fixing
the values of the $n_{\sigma}$. With this respect, points where Stokes
phenomena occur act as borders.
In our simple example (Figure \ref{fig:phi40d}): on one side of the border
(left panel: one single thimble is relevant) we have no $n_{\sigma}$
around; on the other side (right panel: three thimbles) we need to fix the
correct values of three $n_{\sigma}$. 
A key point is that {\em discontinuity in the thimble
decomposition does not imply a discontinuity in a physical
observable} we can be interested in. Therefore, one way of fixing the
$n_{\sigma}$ values is to ensure continuity of an observable 
(or possibly more than one). The interested reader is
referred to \cite{thimbleRMT} for a discussion in the case of the 
$0$-dim $\phi^4$ toy model. \\
The determination of the $n_{\sigma}$ is not the
only point we want to make, nor the most important one. The main message
we want to deliver is that
\begin{itemize} 
\item across points where Stokes phenomena occur, we generally have 
discontinuity of thimble decompositions and continuity of physical
observables and {\em this continuity can build a bridge over different regions};
\item the natural candidates to build the bridge are {\em Taylor expansions}.
\end{itemize}

\section{Taylor expansions on thimbles}
\label{sec:TaylorOnThimbles}

To circumvent multiple thimbles simulations by computing Taylor
expansions we want to
\begin{enumerate}
\item find at least two points where one single thimble contributes to the result,
  either strictly speaking or {\em de facto}, and compute Taylor
  expansions at those points;
\item bridge the gap in between the regions where the previous computations are
  performed: in principle one could show that the Taylor
  expansions join smoothly; in practice, at a given order of the
  expansions, Pad\'e approximants can do (much) better;
\item show that the Pad\'e approximants provide a fairly good control on
  the singularity structure in the complex plane; first of all, this can
  confirm the correctness of the procedure (remember the big question:
  do the radii of convergence enable the analytic continuation we have
  been trying?); also, this can possibly provide {\em in se} extra
  insight into the theory at hand.
\end{enumerate}
\noindent
The approach is based on Taylor expansions (which we eventually trade
for Pad\'e approximants) and those are expansions in 
$\frac{\mu}{X}$, where $X$ is a dimensionful parameter, {\em i.e.}
a mass $m$ or the temperature $T$. It is
therefore clear that we will be blind to any non-perturbative effect
in $\frac{\mu}{X}$. We stress that the latter are not the non-perturbative
effects we are most often concerned with (\ie those in the coupling
constant).\\

To show that our program can indeed be accomplished we present two examples. Although
simple, we will argue that they display general enough features to
support the hope that the method has a potential for further
applications. \\

{\bf $1$-dim Thirring model} We can now fill the gap that was
originally pointed out in \cite{ThirringKiku,PauloAndrei1}. 
The action of the $1$-dim Thirring model is
$$
S = \beta \sum_n (1-\cos(x_n)) - \log(\det D) \;\;
\det D = \frac{1}{2^{L-1}} \left( \cosh(L\hat\mu+i\sum_nx_n) + \cosh(L
  \sinh^{-1}(\hat m))\right)
$$
The chemical potential $\hat\mu$ and the mass $\hat m$ are given
in lattice units. We work at fixed value of mass $m$ (and of $\beta$) 
and there is a single parameter controlling the sign problem, namely
$\mu$. We can obtain a dimensionless quantity by taking the ratio  
$\frac{\mu}{m}=\frac{\hat{\mu}}{\hat{m}}$. Since the
analytic result is known, the single thimble approximation was shown
not to account for the correct result on the entire
$\frac{\mu}{m}$ axis. In our new approach the problem is solved and 
in Figure \ref{fig:thirring} we display the
essential features of our results: as an example, we show results for 
the chiral condensate $\langle \bar\chi \chi \rangle$ (parameters are 
$L=8$, $\beta=1$, $m=2$). We can argue that all the
requirements of the program that we sketched above can be met. There
is a preliminary point we have to make. For real $\beta$ a Stokes
phenomenon is potentially present up to a given value of
$\frac{\mu}{m}$: this involves the dominant thimble $p_{\sigma_0}$ and another
critical point. We denote the latter $p_{\sigma_{\bar 0}}$, following 
the notation of \cite{StudyThirringKiku}. The problem can be easily 
solved by adding a small imaginary part to $\beta$: in this way a Stokes phenomenon
does not take place, a thimble decomposition is in place and while
$p_{\sigma_{\bar 0}}$ could in principle give a contribution to the
result, this is {\em de facto} negligible due to the huge difference
$S_R(p_{\sigma_{\bar 0}})>>S_R(p_{\sigma_0})$. This solves the problem
and any further reference to this point will be omitted in the following.
\begin{enumerate}
\item A first value of $\frac{\mu}{m}$ for which only the dominant
  thimble $p_{\sigma_0}$ accounts for the correct result can be found in a
  very fundamental, yet simple way. The range of values $S_I$ can take 
  on the real axis depends on the values of $\hat{\mu}$ and $\hat{m}$
  and, below a given value of $\frac{\mu}{m}$, this range is limited. 
  By explicit computation of the $S_I^{(\sigma)}(\frac{\mu}{m})$ we
  can show that no unstable thimble associated to a critical
  point $p_\sigma$ other that the dominant one can intersect the
  original domain of integration below a given value
  $\frac{\mu_0}{m}$\footnote{The value of $\hat{m}$ is held fixed.}. 
  Thus for $\frac{\mu}{m}<\frac{\mu_0}{m}$ we can
  easily select a first point at which the dominant thimble provides
  the only contribution to the result. 
  We picked $\frac{\mu}{m}=0.4$ and computed the
  Taylor expansion up to the second derivative.\\
  We now need to find a second value of $\frac{\mu}{m}$ at which the 
  dominant thimble accounts for the complete result and compute the Taylor
expansion on it. In principle we could study the
crossing mechanism between the different curves
$S_I^{(\sigma)}(\frac{\mu}{m})$ (see subsection \ref{sec:Stokes}). In
practice there is a much simpler way to proceed. First of all, we 
point out that the asymptotic value of $\langle \bar\chi \chi
\rangle$ is known: for large enough values of $\mu$ the chiral
condensate is zero. We notice that for $\frac{\mu}{m}=1.4$ the value
of $\langle \bar\chi \chi \rangle$ computed on the dominant thimble is
very close to zero. By inspecting the values of $S_R(p_\sigma)$ for
thimbles other than the fundamental one, we find that, for $\frac{\mu}{m}=1.4$,
$S_R(p_\sigma)>>S_R(p_{\sigma_0})$ for all the critical points but
three, that we denote $\sigma_1$, $\sigma_{\bar 1}$, $\sigma_{\bar 2}$\footnote{We once again adhere to the
  notation of \cite{StudyThirringKiku}.}. Two of them ($\sigma_{\bar
  1}$ and $\sigma_{\bar 2}$) have values of the real action
which are lower than $S_{\mbox{\tiny {min}}}$, which is the minimum value
$S_R$ takes on the original domain of integration: because of this,
the unstable thimbles associated to them can't intersect the original
domain of integration. As for $\sigma_1$, in this simple model it 
does not take that much to show that the unstable thimble attached to 
it does not intersect the original domain of integration (see
the left panel of Figure \ref{fig:thirring}). We conclude that
the dominant thimble $\sigma_0$ can account for the complete result at
this value of $\frac{\mu}{m}$. We have thus selected the
second point we were looking for; at this point the series has been 
computed up to the fifth derivative. One might object that we made use
of the explicit query for intersections between the original domain of
integration and a given unstable thimble, which thing is quite hard to
do in a less simple theory. In the second example we will proceed in a
different way: in principle one could follow the same approach also in this
case\footnote{We will see that we need to have a known value (which
one can trust as correct) and reconstruct the latter by our Taylor
expansion. In this case the asymptotic value at large enough values of
$\frac{\mu}{m}$ is the natural candidate (possibly to be reached in a
two/three steps procedure).}. 
 \item In order to bridge the gap in between the two values of 
$\frac{\mu}{m}$ at which Taylor expansions on the dominant thimble
have been
  computed, one could try to show that the two Taylor expansions do 
  smoothly join. This would actually ask for computing quite a large
  number of derivatives at the lower value of the chemical potential
  (as one can see, the curve is quite flat nearby). As we have already
  pointed out, a Pad\'e approximant can do better. 
  In the middle panel of Figure \ref{fig:thirring} we plot the
  interpolation we got from a Pad\'e approximant on top of the
  analytic result. In order to appreciate how this solves the problem 
  of the inconsistency of single thimble computations we refer the 
  reader to the figures in \cite{ThirringKiku}.    

\begin{figure}[ht]
\begin{center}
\hspace{-0.7cm}
\includegraphics[height=3.75cm,clip=true]{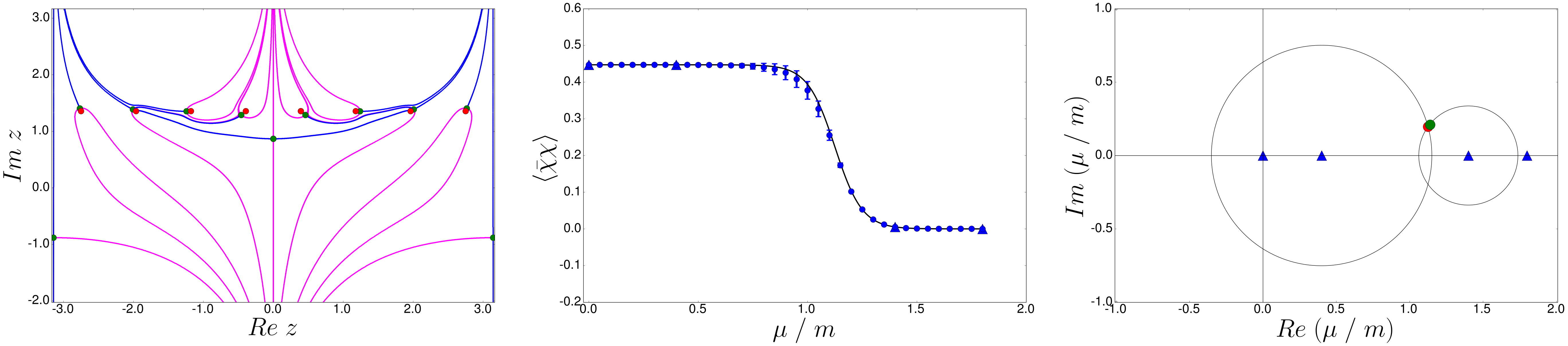}
\end{center}
  \caption{(Left panel) The flow lines highlighting the thimbles structure of the
    $1$-dim Thirring model at $\frac{\mu}{m}=1.4$: stable thimbles are
    depicted in blue, unstable thimbles in magenta. The dominant
    thimble is associated to the critical point sitting at $\Re(z) =
    0$. The critical point $\sigma_1$ is the closest to the latter to
    the right (there is a mirror image to the left as well): 
    notice that the unstable thimble associated to it does
    not intersect the original domain of integration (which is on the
    real axis). (Center panel) The chiral condensate as obtained from
    the analytic solution (continuous black line) and from our Pad\'e
    approximant (we plot points instead of a continuum line so that
    the size of errors are easier to spot.). The points providing
    input to the evaluation of Pad\'e 
    are marked as triangles. (Right panel) Singularity of the solution
  in the complex plane: red point computed from the analytic
  solution, green point is the only pole of our Pad\'e approximant. We
  plot the radii of convergence which are relevant for the expansions
  at hand: our analytic continuation indeed stands on firm ground.}
  \label{fig:thirring}
\end{figure}

\item From the middle panel of Figure \ref{fig:thirring} one can see 
that actually four points were taken into account in our Pad\'e 
procedure (see the four triangles in the plot). On top of the two points
  we discussed previously, two other points enter 
  and act as extra constraints: the values of the condensate at
  $\frac{\mu}{m}=0$ and for $\frac{\mu}{m}$ high enough are known and
  thus they can be taken into account\footnote{Notice that at
    $\frac{\mu}{m}=0$ there is no sign problem and thus one could even
  quite easily compute a high order Taylor expansion.}. 
     The right panel in the figure provides us with a confirmation that the overall
     procedure is under good control. Since the analytic solution is
     known, we know that the latter displays a singularity in the 
complex plane. As one can see, our Pad\'e approximant study captures 
it quite well: see the green and red points practically on top of each other. The
detection of the (expected) singularity is indeed very stable with respect 
to variations in the number of orders that we take into account.
One can thus inspect the convergence radii of our Taylor expansions
and it is indeed confirmed that what we got is a legitimate analytic 
continuation. This is not the only relevant point. The knowledge
of the analytic structure of the solution provides {\em in se} extra
insight into the theory at hand. Needless to say, this is quite often 
one of the main piece of information we will be interested in.
\end{enumerate}

The present computation is essentially only a proof of concept; the
extraction of continuum limit on a line of constant physics has been
obtained as well and will be reported elsewhere \cite{thirringNEW}.

{\bf (The simplest version of) Heavy Dense QCD} In subsection \ref{sec:multiThimble}
we mentioned the multiple thimbles simulation of Heavy Dense
QCD. We now go back to that theory via the application of the Taylor expansion method we
suggest in this paper; the interested reader can compare results with
what we discussed in \cite{HDqcd}. We are dealing with the effective
formulation that can be obtained from QCD by a
combined strong-coupling and hopping parameter expansion. 
One ends up with a 3d effective theory, whose only degrees of freedom are
Polyakov loops \cite{philipsen2,philipsen1,philipsen3}. We tackle
the simplest version of the theory, described by the
action\footnote{Also in this case, we use the hat notation for lattice
 (dimensionless) quantities.}
$$S = S_G + S_F = -\lambda \sum_{<x,y>} \left(\Tr W_x \Tr W_y^\dagger
  + \Tr W_x^\dagger \, \Tr W_y\right)$$
$$     \;\;\;\;\;\;\;\;\;\;\;\;\;\;\;\;\;\;\;\;\;\;\;\;\;\;\;\;\; 
- 2 \sum_x \ln \left( 1 + h_1 \Tr W_x + h_1^2 \, \Tr W_x^\dagger + h_1^3 \right).$$
Here $\lambda = u^{N_t} e^{N_t \cdot
  (4u^4+12u^5-14u^6-36u^7+\ldots)}$, 
$h_1 = (2 k e^{\hat \mu})^{N_t}$, 
$u \approx \frac{\beta}{18}$, $k$ is the hopping parameter 
and $W_x = \prod_{t=1}^{N_t} U_0(x, t)$ is the Polyakov loop.
The theory can be refined by adding to the fermionic part 
a sum extended over nearest neighbours, which couples degrees 
of freedom sitting at different lattice points. The truncation 
at hand amounts to neglecting $O(k^2)$ terms in 
the hopping parameter expansion.
Moreover we are concerned with the cold regime,
where $N_t \gg 1$, $\lambda \approx 0$; in these conditions 
the gauge term of the action is negligible (and is indeed neglected). 
With no interaction among different degrees of freedom $W_x$, one can
write the sum over the degrees of freedom in the action simply as
$\sum_{x=1}^L \ln (1+\dots)$; we studied the model with $L=8$\footnote{Notice that
in the limit in which we work, with interaction among
degrees of freedom missing, dimensionality is a somehow odd concept:
$L=8$ can be interpreted as a tiny $2^3$ 3d system (this is the
canonical HDQCD, coming from actual QCD), but this is not the only
possible interpretation. One could think of a $L=8$ 1d system (in which
case one would have started from 1+1 QCD).}.
Despite its simplicity, the theory displays a sign problem. Also, it
displays features that are interesting in our approach: we will have
the chance to do something different from what we did in the
context of the Thirring model. We stress that also in this case there
is a single parameter controlling the sign problem. We work at fixed
values of $T$ and $m$, so that only $\mu$ varies and it  
enters the game through the combination which defines 
$h_1 = (2 k e^{\hat \mu})^{N_t}=e^{-{\frac{(m-\mu)}{T}}}$. Notice that
(being $m$ and $T$ fixed) this parameter can be $h_1<1$ or $h_1>1$
depending on $\mu$ being $\mu<m$ or $\mu>m$. 
Figure \ref{fig:hdqcd} displays our results for the quark
number density (normalised to $1$). We plot
  results versus the ratio $\frac{\mu}{m}$, the value of $m$ being
  fixed by $k = 0.0000887$.
Also in this case, we can argue that 
all the requirements we formulated for the application of our new 
program can be met.
\begin{enumerate}
\item To find two points at which a Taylor expansion can be computed
  on the dominant thimble, we revert this time to a different
  strategy. We notice that at $\frac{\mu}{m}=1$ (the point that sits
  in the middle of the relevant region) there is no sign
  problem at all; thus computing a Taylor expansion poses no problem. We pick
  two points, to the right ($\mu>m$) and to the left ($\mu<m$) at
  which we will compute Taylor expansions on the dominant thimble. We 
  argue that computations on the dominant thimble at
  those points provide us with the complete result by checking that
  the results obtained by Taylor expansions smoothly (we would say
  perfectly actually) join the result we get at $\frac{\mu}{m}=1$ 
  (of whose correctness we are certain). Remember once again
    that we are concerned with analytic contributions and we will be
    blind to any non-perturbative effect in the expansion parameter. 
  As an extra confirmation, in the left panel of Figure
  \ref{fig:hdqcd}  we plot the relative weights of different thimbles
  as computed from the semiclassical ({\em gaussian}) approximation at
  those values of $\frac{\mu}{m}$ (the dominant thimble virtually
  saturates the normalisation; the weight of the least depressed
  thimble - other than the dominant one - is hardly visible in the
  figure). In the center panel of Figure \ref{fig:hdqcd} one can
  inspect the location of the two points we selected (once again, they
  are marked as triangles).
\item While we plot results as a function of $\frac{\mu}{m}$, our
  expansions are not computed in powers of the latter variable. The
  natural parameter for the expansion is $h_1=e^{-\frac{(\mu-m)}{T}}$: we
  actually expand in $h_1$ (for $\mu < m$) and in
  $h_1^{-1}$ (for $\mu > m$). More precisely, we expand up to the
  second derivative with respect to $h_1$ at $\frac{\mu}{m}=0.9995$
  and we supplement as extra constraints the values of the observable
  and its first derivative at
  $\frac{\mu}{m}=0$ and $\frac{\mu}{m}=1$; we expand up to the second 
  derivative with respect to $h_1^{-1}$ at $\frac{\mu}{m}=1.0005$
  and we supplement as extra constraints the values of the observable
  and its first derivative at
  $\frac{\mu}{m}=1$ and at a value of $\frac{\mu}{m}$ large enough 
  (which is fixed by saturation). Also in this case, we plot in Figure
  \ref{fig:hdqcd} (center panel) the result obtained out of a Pad\'e
  approximant. 

\begin{figure}[ht]
\begin{center}
\includegraphics[height=3.75cm,clip=true]{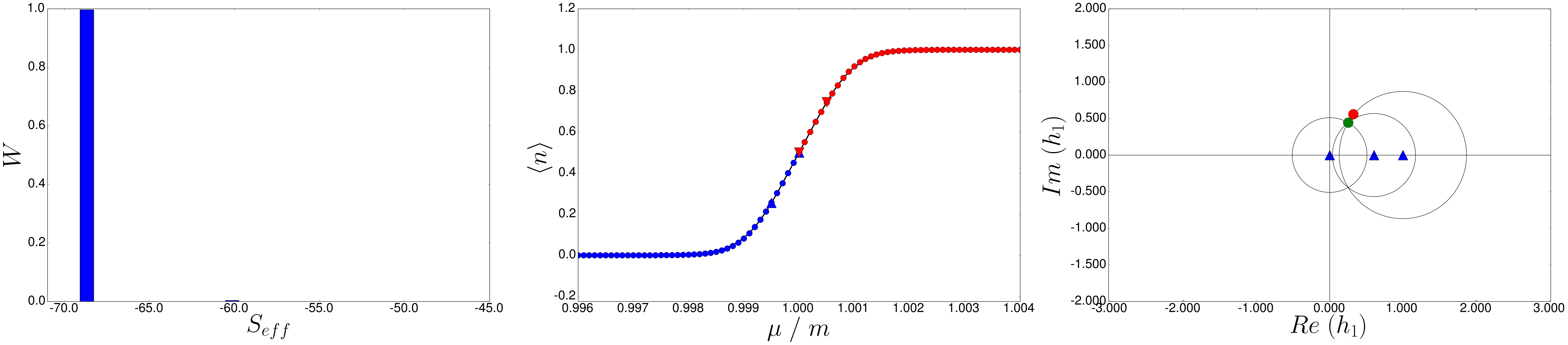}
\end{center}
  \caption{(Left panel) Semiclassical evaluation of the relative
    weight of different thimbles at $\frac{\mu}{m}=0.9995$: weights
    are plotted vs the value of the (real) action. The dominant
    thimble has {\em de facto} weight 1 (it is very hard to detect the
  weight of the thimble with action near the value $-60$). (Center
  panel) The known analytic value for the quark number density is
  correctly reproduced over the entire relevant range of
  $\frac{\mu}{m}$ via the Pad\'e approximant; the points entering the
  computation are marked as triangles (see main text for
  details). (Right panel) The known singularity of the solution as
  analytically known (red point) and as reconstructed by the Pad\'e
  approximant (green point).  }
  \label{fig:hdqcd}
\end{figure}

\item Knowing the analytic solution, also in this case we 
  know of a singularity in the complex plane. In the right panel of 
  Figure \ref{fig:hdqcd} we display how this is
  fairly well reconstructed. The conclusion is once again that the
  procedure we followed is {\em a posteriori} proved correct: the
  convergence radii of the expansions on which we build our
  construction indeed show that the latter is a legitimate
  analytic continuation. 
\end{enumerate}

\section{Conclusions and outlook}
\label{sec:concludendo}

We discussed a new strategy to circumvent multiple thimbles
simulations in the Lefschetz thimble regularisation of a lattice field
theory. The idea is to explore the space of the parameters describing
the theory and find (at least) two points at which the dominant thimble accounts for the
full result: as we saw, there could be different strategies to attain
this. We do expect that in between these two points Stokes phenomena can occur,
so that a non trivial thimble decomposition can be in place. While
Stokes phenomena introduce discontinuities in the thimble decomposition of the
integrals we are interested in, they do not determine in general
discontinuities in physical results. Taylor expansions can thus bridge
the different (disjoint) regions where we can compute on the dominant
thimble only. We are thus aiming at computing the analytic
dependence of our result on the parameter expansion, being blind to
any possible non-perturbative contribution\footnote{We stress once
  again that these are not the {\em standard} non-perturbative effects 
in the coupling}.
Not surprisingly, Pad\'e approximants turn out to be the
most effective tool to implement our program. In particular, they give
us the chance to inspect the analytic structure of our results in the
complex plane and thus the convergence radius of our expansions, so
that our construction can be {\em a posteriori} proved to be a legitimate analytic
continuation. In a quite natural way, computing multiple Taylor
expansions in the complex plane in order to get informations out of
Pad\'e approximants can be a legitimate approach beyond thimbles, no matter
what is the strategy used to obtain the different expansions. In the
end, singularities in the complex plane are quite often one of the
most valuable pieces of information we look for.

\section*{Acknowledgments}
\par\noindent
This work has received funding from the European Union’s Horizon 2020 
research and innovation programme under the Marie Skłodowska-Curie 
grant agreement No. 813942 (EuroPLEx). We also acknowledge support 
from I.N.F.N. under the research project {\sl i.s. QCDLAT}.
This research benefits from the HPC (High Performance Computing) 
facility of the University of Parma, Italy. 

%\newpage

\end{document}